\documentclass[a4paper,11pt]{article}
\pdfoutput=1 

\usepackage{jheppub} 

\usepackage[T1]{fontenc} 

\usepackage[multiple]{footmisc}

\usepackage{graphicx}
\usepackage{amssymb, amsmath,mathtools,amsthm}
\usepackage{hyperref}
\hypersetup{
    colorlinks=true,
    linkcolor=black,          
    citecolor=black,        
    filecolor=black,      
    urlcolor=black          
}

\def\id{{\rm id.}}

\newcommand{\beq}{\begin{equation}}
\newcommand{\eeq}{\end{equation}}

\newcommand{\diff}{{\rm Diff}}
\newcommand{\sdiff}{{\rm Diff_{\rm vol}}}
\newcommand{\vect}{{\rm Vect}}

\newcommand{\lieg}{\mathfrak{g}}
\newcommand{\liegstar}{\mathfrak{g}^*}
\newcommand{\lies}{\mathfrak{s}}

\newcommand{\ad}{{\rm ad}}
\newcommand{\Ad}{{\rm Ad}}
\newcommand{\sch}{{\rm Sch}}
\newcommand{\vir}{\mathfrak{vir}}
\newcommand{\myp}{p}
\newcommand{\bms}{{\rm BMS}_3}
\newcommand{\hr}{\hat{r}}
\newcommand{\hp}{\hat{\phi}}
\newcommand{\be}{\mathbf{e}}

\newcommand{\cJ}{\mathcal{J}}
\newcommand{\cP}{\mathcal{P}}

\title{\boldmath $\bms$ invariant fluid dynamics at null infinity}

\author[a]{Robert F. Penna}
\affiliation[a]{Center for Theoretical Physics, Columbia University,\\
New York, New York 10027, USA}
\emailAdd{rp2835@columbia.edu}

\abstract{

We revisit the boundary dynamics of asymptotically flat, three dimensional gravity.  The boundary is governed by a momentum conservation equation and an energy conservation equation, which we interpret as fluid equations, following the membrane paradigm.  We reformulate the boundary's equations of motion as Hamiltonian flow on the dual of an infinite-dimensional, semi-direct product Lie algebra equipped with a Lie-Poisson bracket.  This gives the analogue for boundary fluid dynamics of the Marsden-Ratiu-Weinstein formulation of the compressible Euler equations on a manifold, $M$, as Hamiltonian flow on the dual of the Lie algebra of $\diff(M)\ltimes C^\infty(M)$.  The Lie group for boundary fluid dynamics turns out to be $\diff(S^1) \ltimes_{\rm Ad} {\rm \vir}$, with central charge $c=3/G$.  This gives a new derivation of the centrally extended, three-dimensional Bondi-van der Burg-Metzner-Sachs ($\bms$) group.  The relationship with fluid dynamics helps to streamline and physically motivate the derivation.  For example, the central charge, $c=3/G$, is simply read off of a fluid equation in much the same way as one reads off a viscosity coefficient.  The perspective presented here may useful for understanding the still mysterious four-dimensional BMS group.

}

\begin{document} 
\maketitle
\flushbottom

\section{Introduction}
\label{sec:intro}

The most common phase spaces in classical mechanics are cotangent bundles.  For example, in $n$ space dimensions, the configuration space of a free particle is $Q=\mathbb{R}^n$ and phase space is $T^*Q\cong \mathbb{R}^{2n}$.  Dynamics is formulated on $T^*Q$ using the canonical Poisson bracket.

One of the next most important phase spaces are the duals of Lie algebras.  Let $G$ be a Lie group, $\lieg$ its Lie algebra, and $\lieg^*$ the dual $\lieg$.  Then $\lieg^*$ is a Poisson manifold with respect to a Lie-Poisson bracket.  There are two Lie-Poisson brackets on $\lieg^*$, which we denote $\{\cdot,\cdot\}_{\pm}$.  They act on functions $F(m),G(m):\liegstar\rightarrow \mathbb{R}$ as
\beq
\{F,G\}_{\pm}(m) = \pm \left\langle m, \left[\frac{\delta F}{\delta m},\frac{\delta G}{\delta m}\right] \right\rangle,\label{eq:lp}
\eeq
where  $\langle\cdot,\cdot \rangle$ is the pairing between $\liegstar$ and $\lieg$.  The functional derivatives $\delta F/\delta m$ and $\delta G/\delta m$ are elements of $\lieg$ and $[\cdot,\cdot]$ is the Lie bracket on $\lieg$.
Define $\liegstar_{\pm}$ to be $\liegstar$ equipped with the $\{\cdot,\cdot\}_{\pm}$ Lie-Poisson bracket.

The latter phase spaces play a key role in fluid dynamics.  The incompressible Euler equations, for a fluid on a manifold $M$, are a Hamiltonian flow on $\liegstar_+$, where $G=\sdiff(M)$ is the group of volume-preserving diffeomorphisms of $M$ \cite{arnold1966geometrie,arnold1969hamiltonian}.   The isentropic, compressible Euler equations are a Hamiltonian flow on the dual of the Lie algebra of $\diff(M)\ltimes C^{\infty}(M)$ \cite{marsden1984semidirect,marsden1984reduction}.  The first factor encodes the fluid's momentum density and the second factor encodes its mass density.  Many further generalizations are known.  For example, fluids carrying conserved charges, magnetized fluids, and nonisentropic fluids can all be treated by adjusting $G$ and the Hamiltonian on $\liegstar_+$ \cite{marsden1984semidirect}.

The rich geometry of the Hamiltonian formulation can be a powerful tool for understanding symmetry, conservation laws, and quantization.  $G$ has a natural action on $\liegstar$ given by the coadjoint action.  The orbits of this action foliate $\liegstar$.  The coadjoint orbits are themselves symplectic manifolds.  
Hamiltonian flows on $\liegstar$ are confined to coadjoint orbits.  According to Kirillov's orbit method, the coadjoint orbits of $\liegstar$ correspond to representations of $G$ \cite{kirillov2004lectures}.  So formulating classical systems on $\liegstar$ gives insights into their quantization.  Now it is a bit unusual to consider quantizing ordinary fluid dynamics\footnote{Although not unheard of: see \cite{2011JHEP...04..102E,Crossley:2015evo}.}, but it is a very interesting prospect for the fluids we are going to describe in this paper, which describe the dual dynamics of gravitational systems.

The dual fluid description of gravity is called the membrane paradigm\footnote{It has a close cousin called fluid/gravity duality \cite{Hubeny:2011hd} but this is not what we will use in this paper.} \cite{1986bhmp.book.....T,1998PhRvD..58f4011P}.  It traces back to Damour's observation that the equations governing the evolution of event horizons can be recast as fluid equations \cite{Damour:1979wya}.  As for ordinary compressible fluids, there is a momentum conservation equation and an energy conservation equation.  These are sometimes called the (Damour-)Navier-Stokes equation and the continuity equation, respectively.  The details of the black hole membrane paradigm depend only on the null character of the horizon, so the formalism can be applied with minor modifications to any null surface.  We will apply the formalism at null infinity of asymptotically flat spacetime in three spacetime dimensions.  We work at null infinity because this is the natural place to define conserved charges in general relativity.  Asymptotic flatness ensures that null infinity is a null surface.

Our goal is to recast the fluid equations governing null infinity as Hamiltonian flow on the dual of a Lie algebra.  Given the close similarity between boundary fluid dynamics and ordinary compressible fluid dynamics, we seek a semidirect product group,
\beq
S=G\ltimes_\Phi V,
\eeq
with $G$ a Lie group, $V$ a vector space, and $\Phi$ a left representation of $G$ on $V$, such that  boundary fluid dynamics is a Hamiltonian flow on $\lies^*_+$ (where $\lies^*_+$ is the dual of the Lie algebra of $S$, equipped with the $+$ Lie-Poisson bracket).  In other words, our task is to use the boundary fluid's equations of motion to identify four pieces of data: $G$, $V$, $\Phi$, and the Hamiltonian.  The solution to this problem is the main result of this paper.

Let us review how this works for the ordinary compressible Euler equations \cite{marsden1984semidirect}.  The continuity equation is
\beq
\partial_t \rho + \nabla \cdot (\rho v) = 0,\label{eq:cont1}
\eeq
where $\rho$ is the fluid's mass density and $v$ is its Eulerian velocity.  Let $\eta_t:M\rightarrow M$ be a map from Lagrangian coordinates, $X$, to Eulerian coordinates, $x$.  It sends the initial positions of the fluid parcels to their positions at time $t$.  It is related to the Eulerian velocity by $\partial_t \eta_t(X)=v(t,\eta_t(X))$.  Now the continuity equation is equivalent to
\beq
\frac{\partial}{\partial t}(\eta_t)^* [\rho(t,x)\thinspace d^nx] = 0,\label{eq:cont2}
\eeq
which says that $\rho$ is simply Lie dragged along by the flow of $\eta_t$.  It is easy to read off $G$, $V$, and $\Phi$ from \eqref{eq:cont2}.  We identify $G=\diff(M)$ with the space of fluid configurations, $\eta_t$.    We identify\footnote{$\rho\thinspace d^nx$ lives in $V^*$ and $\rho$ is identified with an element of $V=C^{\infty}(M)$ using the pairing between $V$ and $V^*$.}  $V=C^{\infty}(M)$ with the space of fluid densities, $\rho$.  We see that $G$ acts on $V$ by the push-forward representation.  So the semidirect product is $\diff(M)\ltimes_* C^\infty(M)$.

Now consider boundary fluid dynamics at null infinity.  The membrane paradigm gives the equations of motion, as we explain in section \ref{sec:membrane}.  Here we simply summarize the result.  In three spacetime dimensions, null infinity is $\mathbb{R}\times S^1$.  We introduce coordinates, $(u,\phi)$, which are the analogue of the Eulerian coordinates, $(t,x)$, of the previous example.   We also introduce  diffeomorphisms, $\zeta_u:S^1\rightarrow S^1$, which may be interpreted as maps from Eulerian coordinates at time $u$ to Lagrangian coordinates.  The fluid's energy density is denoted $\myp$.   The continuity equation then turns out to be
\beq
\frac{\partial}{\partial u} [(\zeta_u')^2 \myp(u,\zeta) - \frac{c}{24\pi}\sch(\zeta_u)] = 0,\label{eq:contb1}
\eeq
where $c=3/G$, the Schwarzian derivative is 
$\sch(\zeta) = (\zeta' \zeta''' - \tfrac{3}{2}(\zeta'')^2)/(\zeta')^2$,
and primes indicate $\partial_\phi$.  The continuity equation \eqref{eq:contb1} is equivalent to
\beq
\frac{\partial}{\partial u}\Ad^*_{\zeta_u^{-1}} (\myp(u,\phi) d\phi^2,ic) = 0,\label{eq:contb2}
\eeq
where $\Ad^*$ is the coadjoint action of $\diff(S^1)$ on $\vir^*$, the dual of the Virasoro algebra.  The energy density, $(\myp(u,\phi) d\phi^2,ic)$, is an element of $\vir^*$ with central charge $c=3/G$.  This form of the continuity equation is precisely analogous to the second form of the continuity equation \eqref{eq:cont2} of the previous example.  We read off the semidirect product structure in exactly the same way. This gives the first part of our main result:
\beq
S=\diff(S^1) \ltimes_{\rm Ad} {\rm \vir_{c=3/G}}.\label{eq:S}
\eeq

The remaining data we need is the Hamiltonian on $\lies^*_+$.  We will find it in section \ref{sec:H} using the boundary fluid's momentum conservation equation, and we will check that computing the corresponding Hamiltonian flow on $\lies^*_+$ gives back the boundary fluid's equations of motion.

We have emphasized the role of $S$ as the fluid's configuration space.  However, $S$ has a double role to play: it also acts as the fluid's symmetry group.  In ordinary fluid dynamics, this is called ``particle relabeling symmetry'' and it expresses the fact that 
ordinary fluid dynamics depends on the velocities but not the spatial labels of fluid elements \cite{arnold1966geometrie,arnold1969hamiltonian,marsden1984semidirect}.  It is a kind of infinite dimensional generalization of ordinary translation invariance.  It manifests as right invariance of the fluid's Hamiltonian under the action of (a subgroup of) $S$.  The associated moment maps are the fluid's convected momentum density and convected energy density.

We will show in section \ref{sec:charges} that $S$ also acts as a symmetry group for boundary fluid dynamics.  This may be interpreted as ``particle relabeling symmetry'' for the boundary fluid.  Projecting the convected momentum density and convected energy density against a basis of $\lies$ and taking Lie-Poisson brackets gives an infinite dimensional algebra of conserved charges.

The symmetry so obtained turns out to be the same thing as the centrally extended, three-dimensional Bondi-van der Burg-Metzner-Sachs ($\bms$) group \cite{Ashtekar:1996cd,Barnich:2006av,2014JHEP...06..129B,Barnich:2015uva}, with central charge $c=3/G$, and the fluid charge algebra is the centrally extended $\bms$ charge algebra.  So one perspective on our result is that it gives a new derivation of extended $\bms$ symmetry and charges.  In certain respects, our derivation is particularly streamlined and physically transparent.  For example, the correct central charge, $c=3/G$, falls out with relatively little effort.  It is simply read off of a fluid equation, with the same ease one reads off a viscosity or other transport coefficient.

Previous work on boundary dynamics in three-dimensional asymptotically flat gravity and $\bms$-invariant actions has appeared in \cite{Barnich:2012rz,Barnich:2013yka,Barnich:2013jla,SalgadoRebolledo:2015yig,Carlip:2016lnw,Barnich:2017jgw}.  That work has some overlap with the present paper but our emphasis on the relationship to fluid dynamics and the Lie-Poisson bracket, and our use of the membrane paradigm as starting point, are new.  The Lie-Poisson bracket made an appearance in \cite{2014JHEP...06..129B,Barnich:2015uva}, but not its relationship to dynamics on $\lies^*$.   

This paper continues the investigation of the relationship between BMS symmetries and fluid symmetries\footnote{For an effective field theory perspective on this relationship, see \cite{Eling:2016xlx,Eling:2016qvx}.} initiated in \cite{Penna:2015gza,Penna:2017bdn}.  The first paper in the series made the observation that BMS conservation laws in four dimensions are equivalent to membrane paradigm conservation laws.  The second paper turned to the near-horizon regions of four-dimensional black holes, where the semi-direct product is $\diff(S^2)\ltimes C^\infty(S^2)$ \cite{Donnay:2015abr,Donnay:2016ejv}, and the relationship to ordinary fluid dynamics is as close as possible.  The present paper is the first in this series to consider central extensions, and the first to explore semi-direct product groups that have not previously appeared in ordinary fluid dynamics (although \cite{guha2005euler,escher2011euler,cismas2016euler} come very close).  In the future, we plan to return to four dimensions, where establishing the correct definitions for the ``superrotation'' subgroup of $S$ at null infinity \cite{2010PhRvL.105k1103B,Barnich:2010eb,Barnich:2011ct,2014PhRvD..90l4028C,2015JHEP...04..076C}, and its possible (central) extensions \cite{2011JHEP...12..105B}, as well as understanding the symmetry underlying the sub-subleading soft graviton theorem \cite{Cachazo:2014fwa,Campiglia:2016efb}, remain outstanding problems.

\section{Boundary fluid dynamics}
\label{sec:membrane}

Our subject is asymptotically flat, three-dimensional gravity.  A reasonable seeming ansatz for the metric is
\beq
ds^2 = \Theta(u,\phi) du^2 - 2dudr + 2 \Xi(u,\phi) dud\phi + r^2 e^{2\varphi(u,\phi)}d\phi^2.
\eeq
However, this metric is not asymptotically flat: $G_{uu} = O(1)$ at large $r$.  To fix this, we take instead\footnote{In fact, the most general asymptotically flat solution of the three-dimensional Einstein equations is known \cite{Barnich:2010eb}.  However, to keep the discussion self-contained we proceed from our ansatz.  Note that the functions $\Theta$ and $\Xi$ of \cite{Barnich:2010eb} are not quite the same as the functions in \eqref{eq:metric}.  However, they are related by a simple transformation which we give explicitly below.}
\beq
ds^2 = [\Theta(u,\phi) - 2r\partial_u \varphi(u,\phi)]du^2 - 2dudr + 2 \Xi(u,\phi) dud\phi + r^2 e^{2\varphi(u,\phi)}d\phi^2.\label{eq:metric}
\eeq
This metric solves the vacuum Einstein equations at large $r$.   

We have given the metric in null coordinates.  To make contact with the membrane paradigm, it will be useful to have the metric in a timelike frame.  To this end, we introduce a timelike triad, 
\begin{align}
\be^{r} 	&= dr,\notag\\
\be^{t} 	&= du+\be^{r^*},\notag\\
\be^{\phi'} 	&= d\phi - \be^{r^{**}},
\end{align}
where $\be^{r^*}$ and $\be^{r^{**}}$ are to be chosen so as to eliminate $g_{tr}$ and $g_{r\phi'}$ from the metric.  Define $\alpha^2=-\Theta+2r\partial_u \varphi$.  A short calculation gives
\begin{align}
\be^{r^*}	&= \left( \alpha^2 + \frac{\Xi^2}{r^2e^{2\varphi}}\right)^{-1}\be^r,\notag\\
\be^{r^{**}}&= \frac{\Xi}{r^2e^{2\varphi}}\left( \alpha^2 + \frac{\Xi^2}{r^2e^{2\varphi}}\right)^{-1}\be^r.
\end{align}
The metric in the timelike frame is
\beq
ds^2 = -\alpha^2 \be^t \otimes \be^t + 2 \Xi \thinspace \be^t \otimes \be^{\phi'} + \be^r \otimes \be^{r^*} 
	+ r^2 e^{2\varphi} \be^{\phi'} \otimes  \be^{\phi'}.
\eeq
We can improve the timelike frame by making it orthonormal.  Define
\begin{align}
\be^{\hat{t}} 	&= \left(\alpha^2+\frac{\Xi^2}{r^2e^{2\varphi}}\right)^{1/2} \be^t,\notag\\
\be^{\hr}	&= \left( \alpha^2 + \frac{\Xi^2}{r^2e^{2\varphi}}\right)^{-1/2} \be^r,\notag\\
\be^{\hp}	&=  \frac{\Xi}{r e^{\varphi}}\be^{t} + r e^\varphi \be^{\phi'}.
\end{align}
Now the metric is simply
\beq
ds^2 = -\be^{\hat{t}} \otimes \be^{\hat{t}}
	+ \be^{\hr} \otimes \be^{\hr} 
	+ \be^{\hp} \otimes  \be^{\hp}.
\eeq

Fix a surface at large but finite $r=r_0$, with unit outward normal $n=\be^{\hr}$.  This surface plays the role of the ``stretched horizon'' of the black hole membrane paradigm.  We are ultimately interested in the limit $r_0\rightarrow \infty$.  The membrane paradigm assigns ``stretched infinity'' a surface stress-energy tensor, $t_{\mu\nu}$.  It is defined using the Israel junction condition to be the surface stress-energy tensor required to terminate the gravitational field at $r=r_0$.  Introduce the projection operator
\begin{align}
h_{\mu\nu} = g_{\mu\nu} - n_\mu n_\nu.
\end{align}
The extrinsic curvature of stretched infinity is ${K^\mu}_\nu={h^\delta}_\nu \nabla_\delta n^\mu$, and the surface stress-energy tensor is
\beq
t_{\mu\nu} = \frac{1}{8\pi G}(K_{\mu\nu}-K h_{\mu\nu}),
\eeq
where $K={K^\mu}_\mu$.
The vacuum Einstein equations imply the conservation laws
\beq
\sqrt{-h} h_{a\mu}{t^{\mu\nu}}_{|\nu} = 0,
\eeq
where the 2-covariant derivative is ${t^{\mu\nu}}_{|\nu} = {h^\delta}_\nu \nabla_\delta t^{\mu\nu}$.  There are two conservation laws, corresponding to $a=t,\phi$.  They are the energy and momentum equations for the boundary fluid.

Consider the energy equation:
\beq
\sqrt{-h} h_{t\mu}{t^{\mu\nu}}_{|\nu} =
	\frac{\partial}{\partial u}\left[e^{2\varphi}p
	+\frac{c}{24\pi}\left(\frac{1}{2}(\varphi')^2-\varphi''\right)\right]=0,
\eeq
where $p=\Theta/(16\pi G)$ and $c=3/G$.  We bring this into a more familiar form by introducing a time-dependent diffeomorphism, $\zeta_u:S^1\rightarrow S^1$, defined by $\zeta_u'(\phi) = e^{2\varphi(u,\phi)}$, with Schwarzian derivative $\sch(\zeta) = (\zeta' \zeta''' - \tfrac{3}{2}(\zeta'')^2)/(\zeta')^2$.  The energy equation becomes 
\beq
\frac{\partial}{\partial u}\left[(\zeta_u')^2 p(u,\zeta)
	-\frac{c}{24\pi}\sch(\zeta_u)\right]=0.
\eeq
Equivalently, 
\beq
\frac{\partial}{\partial u}\Ad^*_{\zeta_u^{-1}} (\myp(u,\phi) d\phi^2,ic) = 0,\label{eq:energy}
\eeq
where $\Ad^*$ is the coadjoint action\footnote{See Appendix \ref{sec:app}.} of $\diff(S^1)$ on $\vir^*$, the dual of the Virasoro algebra.  We identify the energy density, $(\myp(u,\phi) d\phi^2,ic)$, with an element of $\vir^*$ with central charge $c=3/G$.
As explained in the introduction, it is now straightforward to read off the semi-direct product 
\beq
S=\diff(S^1) \ltimes_\Ad {\rm \vir_{c=3/G}}.
\eeq
For the remainder of this paper, $G=\diff(S^1)$, $V=\vir$, $\Phi=\Ad$, and $S=G\ltimes_\Phi V$.

It remains to find the Hamiltonian on $\lies^*_+$ and recover the boundary fluid equations as Hamiltonian flow, and to understand the interpretation of $S$ as a symmetry group.  This is the objective of the next two sections.

\section{Hamiltonian}
\label{sec:H}

Turn now to the momentum equation:
\beq\label{eq:premomentum}
\sqrt{-h} h_{\phi\mu}{t^{\mu\nu}}_{|\nu} = \partial_u (e^\varphi \tilde{\jmath}) - e^\varphi p' = 0,
\eeq
where $\tilde{\jmath}=\Xi/(8\pi G)$.  We can bring this into a more familiar form by introducing $j=e^{-\varphi}[\tilde{\jmath}-e^{-\varphi}\int d\tilde{u}e^\varphi p'].$  The momentum equation becomes 
\beq
\frac{\partial}{\partial u} \Ad^*_{\zeta_u^{-1}} [j(u,\phi)d\phi^2] = 0,\label{eq:momentum0}
\eeq
with $j(u,\phi)d\phi^2$ understood to be an element of $\lieg^*$.  Let $\xi(u,\phi) = \partial_u \zeta_u(\phi)$.  The infinitesimal form of the momentum equation is
\beq
\partial_u [j(u,\phi)d\phi^2]
	 = \ad^*_{\xi\partial_\phi}[j(u,\phi)d\phi^2] 
	 =  (2j \xi' + j' \xi)d\phi^2.\label{eq:momentum}
\eeq

We identify elements of $\lies^*_+$ with pairs $(j,p)$ (we have suppressed the central charge to ease the notation).  Define $H:\lies_+\rightarrow \mathbb{R}$ as
\beq
H=\int_{S^1} \xi j d\phi.\label{eq:H}
\eeq
This is conserved in the sense that
\beq
\int_{S^1} \xi \partial_u j d\phi
 = \int_{S^1} \xi [2j \xi' + j' \xi] d\phi
 = \int_{S^1} (\xi^2 j)' d\phi = 0.
\eeq
Furthermore, $\delta H/\delta j = \xi$ is the generator of (an infinite dimensional generalization of) ``rotations.''  For these reasons, we identify $H$ with the fluid's Hamiltonian.  We will now verify that the associated Hamiltonian flow on $\lies^*_+$ is equivalent to the boundary fluid's equations of motion.

The Lie bracket on $\lies=\lieg \times_\ad V$ is
\beq
[(X,v),(Y,u)] = ([X,Y],\ad_{X}u - \ad_{Y}v),\label{eq:Lie}
\eeq
where $X,Y\in \lieg$ and $v,u\in V$.
The $+$ Lie-Poisson bracket \eqref{eq:lp} of $F,G:\lies^*\rightarrow \mathbf{R}$ is
\beq
\{F,G\}_+(j,p) = \left\langle j,  \left[\frac{\delta F}{\delta j},\frac{\delta G}{\delta j}\right]  \right\rangle
	+\left\langle p, \ad_{\delta F/\delta j} \frac{\delta G}{\delta p} \right\rangle
	- \left\langle p, \ad_{\delta G/\delta j} \frac{\delta F}{\delta p} \right\rangle.\label{eq:lp2}
\eeq
The Hamiltonian vector field, $X_H$, is defined by
\beq
\langle dF, X_H \rangle = \{F,H\}_+,
\eeq
for all $F:\lies^*\rightarrow \mathbb{R}$.  It is
\beq
X_H(j,p) 
	= - \left(\ad^*_{\delta H /\delta j}\thinspace j + \ad^*_{\delta H/\delta p}\thinspace p ,
		\thinspace \ad^*_{\delta H/\delta j}\thinspace  p  \right),
\eeq
where $\ad_{\delta H/\delta p}:\lieg\rightarrow V$ is defined by  $\ad_{\delta H/\delta p}(\xi) = -\ad_\xi (\delta H/\delta p)$ and  $\ad^*_{\delta H/\delta p}$ is its dual.
With the Hamiltonian given by \eqref{eq:H}, the equations of motion, $\partial_u (j,p)=-X_H(j,p)$, are 
\begin{align}
\partial_u [j(u,\phi)d\phi^2] 	&= \ad^*_{\xi\partial_\phi} [j(u,\phi)d\phi^2],\\ 
\partial_u [(p(u,\phi)d\phi^2,c)]	&= \ad^*_{\xi\partial_\phi} [(p(u,\phi)d\phi^2,c)],
\end{align}
which are equivalent to the boundary fluid's energy \eqref{eq:energy} and momentum \eqref{eq:momentum} equations.

It remains to understand the interpretation of $S$ as a symmetry group.  This is the objective of the next section.

\section{Symmetries and conservation laws}
\label{sec:charges}

Define\footnote{We may further define $\tilde{\Xi}=8\pi G \cJ$ and $\tilde{\Theta}=16\pi G \cP$.  Then $\tilde{\Xi}=\tilde{\Xi}(\phi)$ and $\tilde{\Theta}=\tilde{\Theta}(\phi)$ are independent of $u$.  These functions, $\tilde{\Xi}$ and $\tilde{\Theta}$, are what reference \cite{Barnich:2010eb} denotes $\Xi$ and $\Theta$.}
\begin{align}
\cJ  &= \Ad^*_{\zeta_u^{-1}} [j(u,\phi)d\phi^2],\\
\cP &= \Ad^*_{\zeta_u^{-1}} (\myp(u,\phi) d\phi^2,ic).
\end{align}
The boundary fluid equations of motion \eqref{eq:energy} and \eqref{eq:momentum0} imply $\cJ=\cJ(\phi)$ and $\cP=\cP(\phi)$ are conserved.  

We noted earlier that $\zeta_u$ may be interpreted as a map from Eulerian coordinates at time $u$ to Lagrangian coordinates.  The justification for this terminology comes from comparing the ordinary compressible fluid continuity equation \eqref{eq:cont2} to the boundary fluid continuity equation \eqref{eq:contb2}, and noting that both $\eta_t$ and $\zeta_u^{-1}$ act on the right.  Pushing this nomenclature further, we identify $j$ and $p$ with the boundary fluid's Eulerian momentum density and Eulerian energy density, and we identify $\cJ$ and $\cP$ with the boundary fluid's convected momentum density and convected energy density.  Conservation of $\cJ$ and $\cP$ may then be understood as a consequence of a kind of particle-relabeling symmetry for the boundary fluid.

Consider a Fourier basis for $\lies=\vect(S^1)\times_\ad \vir$ given by
\begin{align}
X_m &= (e^{im\phi}\partial_\phi,(0,0))\\
v_m &= \left(0,\left(e^{im\phi}\partial_\phi,-\frac{i}{24}\delta_m^0\right)\right).
\end{align}
The Lie brackets \eqref{eq:Lie} are
\begin{align}
i[X_m,X_n] &= (m-n)X_{m+n},\notag\\
i[X_m,v_n] &= (m-n)v_{m+m} + \frac{Z}{12}m(m^2-1)\delta^0_{m+n},\notag\\
i[v_m,v_n] &= 0,
\end{align}
with $Z=(0,(0,-i))$.  Define
\begin{align}
\cJ_m & = \int_{S_1} X_m \cdot \cJ, \\
\cP_m & = \int_{S_1} v_m \cdot \cP. 
\end{align}
The Lie-Poisson brackets \eqref{eq:lp2} are 
\begin{align}
i\{\cJ_m,\cJ_n\}_+ &= (m-n)\cJ_{m+n},\notag\\
i\{\cJ_m,\cP_n\}_+ &= (m-n)\cP_{m+n} + \frac{c}{12}m(m^2-1)\delta^0_{m+n},\notag\\
i\{\cP_m,\cP_n\}_+ &= 0,
\end{align}
which is the centrally extended $\bms$ charge algebra \cite{Barnich:2006av,2014JHEP...06..129B,Barnich:2015uva}.

\section{Discussion}
\label{sec:discuss}

Boundary fluid dynamics is governed by two equations involving three unknown functions: $\zeta_u$, $p$, and $j$.  These functions are analogous to the configuration map, $\eta_t$, mass density $\rho$, and momentum density, $\rho v$, of the ordinary compressible Euler equations.  However, there is an important conceptual difference.  In ordinary fluid dynamics, these three functions are not independent: the configuration map determines the fluid velocity (and vice versa) via the relation $v(t,\eta_t(X)) = \partial_t \eta_t(t,X)$.

There is no such relation for the boundary fluid.  $\zeta'_u(\phi)=e^{2\varphi(u,\phi)}$ is an arbitrary function that must be picked at the outset to close the equations of motion.  Once this function is fixed, the equations of motion describe how $p$ and $j$ are advected along by the flow of $\zeta_u$.   This is related to an observation of Carlip \cite{Carlip:2016lnw}, who regards the function $\varphi(u,\phi)$ appearing in the metric \eqref{eq:metric} as an arbitrary choice of vacuum.  One could also describe this as a choice for how one sets up the boundary laboratory at infinity.  Three-dimensional asymptotically flat vacuum Einstein gravity has almost no dynamics; the question we have been led to study is how $p$ and $j$ evolve as one varies the laboratory, $\zeta_u$.  If one chooses the ``trivial laboratory,'' $\zeta_u(\phi) = \phi$ (which corresponds to setting $g_{\phi\phi} = r^2$ in the metric), then $j$ and $p$ are simply constant in time.

Since the function $\varphi(u,\phi)$ appearing in the metric \eqref{eq:metric} is arbitrary, it seems one could choose stronger boundary conditions at infinity to force $\zeta_u$ and $j$ to have the same relation to each other that they enjoy in ordinary fluid dynamics.  This would make the boundary dynamics more natural in some ways.  But it appears to be an unnatural thing to do from the gravity side, which is why we have avoided taking this step in this paper.

On a related note, our Hamiltonian is really a family of Hamiltonians, $H_\xi = \int_{S^1} \xi j d\phi = H$, depending on the choice of $\zeta_u$.  We could enlarge phase space so that $\xi$ becomes a proper phase space variable.  But this extra sector of phase space has no interesting dynamics, so we prefer to quotient it out and study the dynamics on $\lies^*_+$, using $H_\xi$.  Actually, a similar trick appears in ordinary compressible fluid dynamics, where it is sometimes convenient to regard the dynamics as taking place on $T^*\diff(M)$, and use a family of Hamiltonians, $H_{\rho_0}$, depending on an arbitrary choice for the initial mass density, $\rho_0$ \cite{marsden1984semidirect}.

The present discussion helps to clarify the transformation relating $\tilde{\jmath}$ and $j$ that we introduced beneath \eqref{eq:premomentum}.  That transformation involves an integral over all of time.  In ordinary fluid dynamics, one would not normally introduce a field redefinition that involves an integral over all of time, as all of the fields are interdependent.  However, in our setup, $p(u,\phi)$ is fixed by $\zeta_u$ alone, so it is straightforward to solve for $p$ and $j$ and then reconstruct $\tilde{\jmath}$.

Finally, let us comment on a possible generalization of our results.  We have focused on asymptotically flat gravity because in this case the boundary is null and we are as close as possible to the usual membrane paradigm formalism.  However, generalizations to asymptotically (anti-)de Sitter ((A)dS) gravity should be possible.  The definition of the membrane stress-energy tensor and its conservation laws relies only on the Israel junction condition, which can be applied at any hypersurface, null or otherwise.  The null condition is only used in a crucial way to define the shear viscosity and resistivity of the membrane.  These did not appear in the present paper because we are in three spacetime dimensions (so no gravity waves, and consequently no shear viscosity) and we do not consider electrodynamics.  The viscosity and resistivity encode an outgoing boundary condition that is unique to null surfaces (in a certain reference frame, all modes are outgoing).  However, at non-null boundaries of spacetime there are analogous boundary conditions and they can perhaps be encoded in some modification of the usual membrane viscosity and resistivity. The generalization of our results to ${\rm AdS}_3$ would probably lead back to the well-known representation of ${\rm AdS}_3$ boundary dynamics as Liouville theory \cite{Coussaert:1995zp}.  It is also worth noting that the Schwarzian action that describes the boundary dynamics of ${\rm AdS}_2$ \cite{Maldacena:2016upp} has the same form as the membrane paradigm action of \cite{1998PhRvD..58f4011P} (they are both integrals of the trace of the extrinsic curvature of the boundary).  Finally, we note that the membrane paradigm in asymptotically AdS spacetime is closely related to the radial Hamiltonian analysis of the holographic renormalization group (e.g., \cite{deBoer:1999tgo,Martelli:2002sp,Papadimitriou:2004ap,Papadimitriou:2005ii}).

\acknowledgments

We thank Geoffrey Comp\`{e}re for comments, and we are very grateful to Blagoje Oblak for helpful correspondence at the beginning and end of this project and for his close reading of an earlier draft.  This research was supported by a Prize Postdoctoral Fellowship in the Natural Sciences at Columbia University.

\appendix

\section{Coadjoint actions}
\label{sec:app}

Consider a Lie group, $G$, with Lie algebra, $\lieg$, and dual Lie algebra, $\lieg^*$.  The group, $G$, acts on itself via conjugation
\beq
\quad c_g:h \rightarrow ghg^{-1}, \quad g,h\in G.
\eeq
We want to build an action of $G$ on $\lieg$.  Elements of $\lieg$ are tangent vectors to $G$ at the identity.  Given a tangent vector $\xi\in \lieg$, consider the curve $h_t = \id + t\xi + \dots$ in $G$.  The adjoint action of $G$ on $\lieg$ is defined as
\beq\label{eq:Ad}
\Ad_g\thinspace \xi = \frac{d}{dt}\bigg\vert_{t=0} gh_tg^{-1}.
\eeq
Given a second tangent vector, $\eta\in \lieg$, consider the curve $g_t = \id + t\eta + \dots$ in $G$.  This lets us build the adjoint action of $\lieg$ on itself:
\beq
\ad_\eta \thinspace \xi = \frac{d}{dt}\bigg\vert_{t=0} \Ad_{g_t} \thinspace \xi.
\eeq
This is the same thing as the Lie bracket: $\ad_\eta \thinspace\xi = [\eta,\xi]$.

Elements of $\lieg^*$ are functions on $\lieg$. So we have a pairing, $\langle \cdot,\cdot \rangle$, between $\lieg$ and $\liegstar$ given by $\langle\mu,\xi\rangle=\mu(\xi)$, where $\mu\in \liegstar$ and $\xi\in\lieg$.  The coadjoint action of $G$ on $\liegstar$ is defined by
\beq\label{eq:Adstar}
\langle \Ad^*_g\thinspace \mu , \xi \rangle = \langle \mu,\Ad_{g^{-1}}\thinspace \xi \rangle,
\eeq
for all $\xi\in \lieg$.  The coadjoint action of $\lieg$ on $\liegstar$ is defined by
\beq\label{eq:adstardef}
\langle \ad^*_\eta\thinspace \mu , \xi \rangle = -\langle \mu,\ad_\eta\thinspace \xi \rangle,
\eeq
for all $\xi\in \lieg$.

\subsection{Virasoro}

Our goal is to work out the coadjoint action of $\diff(S^1)$ on the dual of the Virasoro algebra.  $\diff(S^1)$ is the group of smooth diffeomorphisms of the circle.  This is an infinite dimensional Lie group with multiplication given by composition.  Its Lie algebra, $\vect(S^1)$, is the algebra of smooth vector fields on the circle.  The Lie bracket is minus the usual vector field commutator.  We identify the dual of $\vect(S^1)$ with the space of quadratic differentials on the circle.  The pairing is
\beq
\langle f(\phi) \partial_\phi, u(\phi) d\phi^2 ,  \rangle = \int_{S^1} f(\phi)u(\phi) d\phi.
\eeq

The Virasoro algebra, $\vir$, is a central extension of $\vect(S^1)$.  It is an infinite dimensional Lie algebra of the form $\vect(S^1)\times i\mathbb{R}$.  The Lie bracket on $\vir$ is
\beq\label{eq:virbracket}
[(f\partial_\phi,-ia),(g\partial_\phi,-ib)] 
	= \left([f\partial_\phi,g\partial_\phi],-\frac{i}{48\pi}\omega(f\partial_\phi,g\partial_\phi)\right),
\eeq
where the bracket on the rhs is the Lie bracket on $\vect(S^1)$.  The map $\omega:\vect(S^1)\times\vect(S^1)\rightarrow \mathbb{R}$ is the Gelfand-Fuchs cocycle:
\beq
\omega(f\partial_\phi,g\partial_\phi) = 2 \int_{S^1} f'(\phi) g''(\phi)d\phi.
\eeq
Elements of the dual, $\vir^*$, may be identified with pairs $(f(\phi)d\phi^2,ia)$.  The pairing between $\vir$ and $\vir^*$ is
\beq
\langle (f\partial_\phi,-ia) , (ud\phi^2,ib) \rangle = \int_{S^1} fu d\phi+ ab.
\eeq

The adjoint action of $\vir$ on itself is simply given by \eqref{eq:virbracket}:
\beq
\ad_{(f\partial_\phi,-ia)} (g\partial_\phi,-ib) 
	=  \left([f\partial_\phi,g\partial_\phi],-\frac{i}{48\pi}\omega(f\partial_\phi,g\partial_\phi)\right).
\eeq
Evaluating the coadjoint action of $\vir$ on $\vir^*$ requires a bit more work.  First, recall the definition \eqref{eq:adstardef}:
\beq
\langle \ad^*_{(f\partial_\phi,-ia)}\thinspace (ud\phi^2,ic), (g\partial_\phi,-ib) \rangle 
	= -\langle (u  d\phi^2,ic),\ad_{(f\partial_\phi,-ia)}\thinspace (g\partial_\phi,-ib) \rangle,
\eeq
for all $(g\partial_\phi,ib) \in \vir$.  Evaluating the rhs and rearranging gives:
\beq\label{eq:viradstar}
\ad^*_{(f\partial_\phi,-ia)}\thinspace (ud\phi^2,ic) 
	= -\left(\left(u'f +2uf'-\frac{c}{24\pi}f'''\right)d\phi^2,0\right).
\eeq

The coadjoint action of $\diff(S^1)$ on $\vir^*$ is
\beq\label{eq:virAdstar}
\Ad^*_{\zeta^{-1}} (u d\phi^2,ic) 
 =  \left(\left((\zeta')^2 u(\zeta)
	-\frac{c}{24\pi}\sch(\zeta)\right)d\phi^2,ic\right),
\eeq
where $\zeta\in \diff(S^1)$ and the Schwarzian derivative is $\sch(\zeta) = (\zeta' \zeta''' - \tfrac{3}{2}(\zeta'')^2)/(\zeta')^2$.  Rather than deriving \eqref{eq:virAdstar} directly using \eqref{eq:Ad} and \eqref{eq:Adstar}, we will check that the infinitesimal version of \eqref{eq:virAdstar} is given by \eqref{eq:viradstar}.  Let
\begin{align}
\zeta_t(\phi) &= \phi + t f(\phi) + \dots\\
\zeta_t(\phi)^{-1} &= \phi-tf(\phi)+\dots
\end{align}
Note
\beq
\frac{d}{dt}\bigg\vert_{t=0}\sch(\zeta_t)
	= \frac{d}{dt}\bigg\vert_{t=0} \frac{(1+tf')tf'''-\tfrac{3}{2}(tf'')^2}{(1+tf')^2}
	= f'''.
\eeq
Now it is straightforward to check that 
\beq
\ad^*_{(f\partial_\phi,-ia)}\thinspace (ud\phi^2,ic) 
	=  \frac{d}{dt}\bigg\vert_{t=0} \Ad^*_{\zeta_t} (u d\phi^2,ic).
\eeq

\bibliography{ms}

\end{document}